\documentclass[twocolumn, notitlepage, superscriptaddress, aps]{revtex4-1}
\usepackage{xcolor}
\usepackage{amsmath}
\usepackage{amsthm}
\usepackage{graphicx}
\usepackage{float}
\usepackage{hyperref}
\usepackage{ulem}
\usepackage{caption}
\usepackage[position=top]{subfig}
\usepackage{framed}
\newcommand{\omg}{\omega}
\usepackage{color}
\newcommand{\rev}[1]{#1}

\begin{document}
\title{Compact All-Fiber Quantum-Inspired LiDAR with \(>\) 100dB Noise Rejection and Single Photon Sensitivity}
\author{Han Liu}
\affiliation{The Edward S.~Rogers Department of Electrical and Computer Engineering, University of Toronto, 10 King's College Road, Toronto, Ontario M5S 3G4, Canada}
\author{Changhao Qin}
\affiliation{The Edward S.~Rogers Department of Electrical and Computer Engineering, University of Toronto, 10 King's College Road, Toronto, Ontario M5S 3G4, Canada}
\author{Georgios Papangelakis}
\affiliation{The Edward S.~Rogers Department of Electrical and Computer Engineering, University of Toronto, 10 King's College Road, Toronto, Ontario M5S 3G4, Canada}
\author{Meng Lon Iu}
\affiliation{The Edward S.~Rogers Department of Electrical and Computer Engineering, University of Toronto, 10 King's College Road, Toronto, Ontario M5S 3G4, Canada}
\author{Amr S.~Helmy}
\email{a.helmy@utoronto.ca}
\affiliation{The Edward S.~Rogers Department of Electrical and Computer Engineering, University of Toronto, 10 King's College Road, Toronto, Ontario M5S 3G4, Canada}
\begin{abstract}
	Entanglement and correlation of quantum light can enhance LiDAR sensitivity in the presence of strong background noise.
However, the power of such quantum sources is fundamentally limited to a stream of single photons and cannot compete with the detection range of high-power classical LiDAR transmitters.
To circumvent this, we develop and demonstrate a quantum-inspired LiDAR prototype based on coherent measurement of classical time-frequency correlations.
This system uses a high-power classical source and maintains the high noise rejection advantage of quantum LiDARs.
In particular, we show that it can achieve over 100dB rejection (with 100ms integration time) of indistinguishable \rev{(with statistically identical properties in every degrees of freedom)} in-band noise while still being sensitive to single photon signals.
In addition to the LiDAR demonstration, we also discuss the potential of the proposed LiDAR receiver for quantum information applications.
In particular, we propose the chaotic quantum frequency conversion technique for coherent manipulation of high dimensional quantum states of light.
It is shown that this technique can provide improved performance in terms of selectivity and efficiency as compared to pulse-based quantum frequency conversion.
\end{abstract}
\maketitle
\section{introduction}
In any optical sensing instrumentation, the light source and detection system used play a pivotal role in dictating the performance.
In recent years, a radical approach to enhancing optical sensing system sensitivity has been to use quantum light sources and measure their non-classical properties.
This serves to surpass the performance limit imposed by the classical laws of physics.
A manifestation of this idea in the target detection domain is quantum illumination (QI), where quantum entanglement is utilized to reject the background noise of the target detection channel\cite{lloyd2008enhanced,zhang2015entanglement,barzanjeh2015microwave,liu2020joint}.
In a QI setup, the probe light that is entangled with, the locally stored reference light, interrogates the target.
Back-reflected probe light (if the target is present) mixed with strong background noise light is collected and undergoes a joint detection measurement along with the reference light to determine the target's presence or absence.
In contrast to the common perception of quantum light being fragile, \rev{the performance advantage of QI over classical detection is most pronounced in the high loss and high noise regime.} Similar enhancement of LiDAR sensitivity has also been demonstrated through phase-insensitive measurement of photon-photon correlations\cite{liu2019enhancing,zhang2020multidimensional,blakey2022quantum}.\\

Despite its unrivaled performance over classical LiDAR with equal probe power, practical applications of QI are severely curtailed owing, in part to its fundamental power limit: not only the flux of an entangled light source is difficult to increase, but also the performance enhancement it offers diminishes as the power increases\cite{tan2008quantum,sanz2017quantum,nair2020fundamental}.
As such, it is difficult for QI to meet the demand of real-world sensing applications where high probe power is needed to extend the detection range beyond that of a laboratory setup.
Therefore, a natural line of inquiry could pose the question of whether it is possible to borrow methodology from QI to enhance classical LiDAR protocols while retaining the essential performance enhancement.
Similar approaches, other than QI have already been proven successful for sensing protocols that were initially believed to rely on quantum effects, those include ghost imaging\cite{shapiro2008computational} and quantum optical coherence tomography\cite{kaltenbaek2008quantum,lavoie2009quantum}.
In this work, we demonstrate a LiDAR design and the associated prototype with a similar setup and operation principle as a QI, except that it uses a classical time-frequency correlation source with high power thereby greatly enhancing the potential for achieving a significant operating range.
It is shown that by using classical time-frequency correlation, probe light that has random and chaotic time-frequency characteristics can be selectively converted to a single frequency with near-unity quantum efficiency while in-band noise \rev{with identical time-frequency characteristics} can be reduced to a negligible level.\\

In addition, it is interesting to note that this noise rejection technique is conceptually related to quantum frequency conversion (QFC) \cite{eckstein2011quantum,reddy2013temporal,huang2013mode,mckinstrie2012quantum,brecht2015photon}, a quantum information processing technique in which a particular time-frequency mode (probe) is selectively separated from the rest of the band (indistinguishable noise) with preserved quantum properties.
For this reason, we term the LiDAR protocol as chaotic-QFC LiDAR.
Nevertheless, the proposed receiver is different from existing QFC protocols in its use of chaotic time-frequency modes in which the probe light resides.
It can be shown that the high-dimensional and chaotic nature of chaotic modes can provide substantially improved performance in terms of efficiency and selectivity as compared to conventional Hermite-Gaussian modes based QFC.

The rest of this manuscript is organized as follows.
In the first section, we formulate the theory of chaotic-QFC and confirm its validity with numerical Monte Carlo simulation.
In the result section, we design and build a LiDAR prototype that resembles our theoretical formulation.
We then experimentally benchmark this setup in terms of different LiDAR performance metrics including quantum efficiency, noise resilience, and ranging accuracy, which are in good agreement with our theoretical prediction.
In the discussion section, we draw connections and differences between the classical time-frequency correlation used in this work and non-classical time-frequency entanglement that can be used in QI\cite{liu2020joint}, as well as other correlation-based LiDAR protocols.
Finally, we extend the theoretical framework of chaotic-QFC to showcase its application in quantum information applications.

\newcommand{\avg}[1]{\langle #1 \rangle}
\section{theory}
\begin{figure*}[t]
	\centering
	\includegraphics[width=1.8\columnwidth]{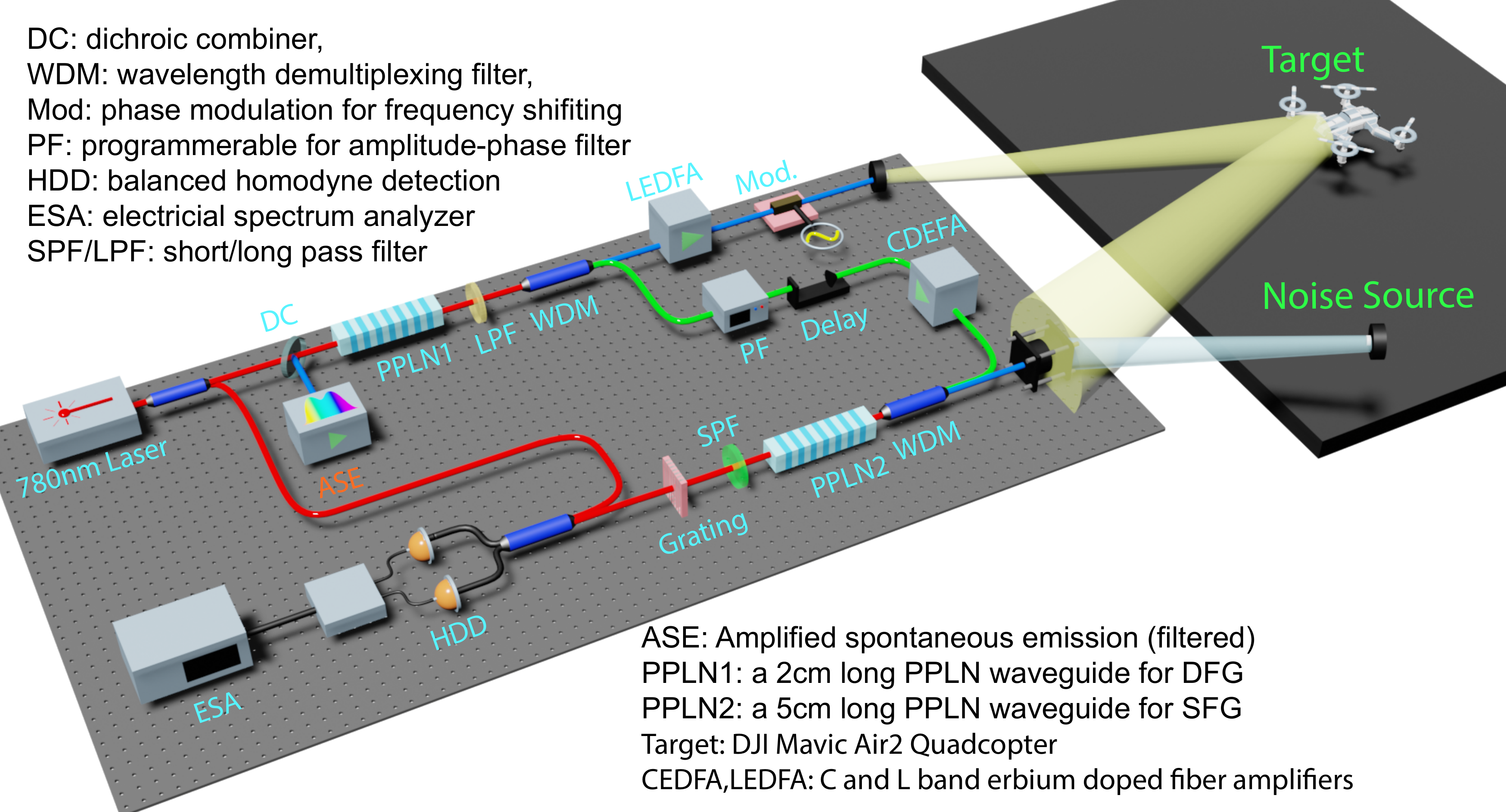}
	\caption{The experimental setup of the LiDAR system.
\label{setup}}
\end{figure*}

In this section, we theoretically model the chaotic-QFC LiDAR that is based on classical time-frequency correlation.
In particular, we analyze the generation and detection process of such correlation and show how it enables substantial noise suppression while still maintaining single photon sensitivity for LiDAR operation.\\  
\begin{figure*}[t]
	\centering
	\includegraphics[width=1.9\columnwidth]{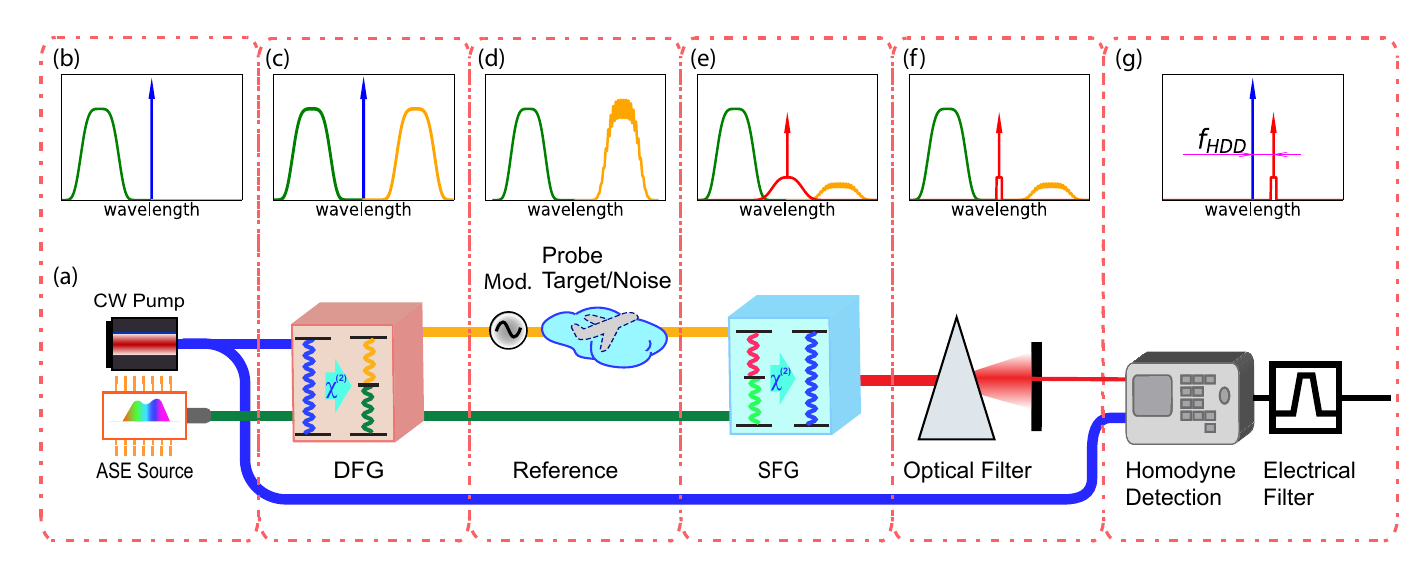}\\
	\caption{(a)The schematic diagram of the chaotic-QFC LiDAR and its various stages.
(b) the spectrum of the DFG pump (blue) light and ASE (green) before the DFG process, (c) the spectrum of the probe (yellow), reference (green), and DFG pump light after the DFG process, (d) the spectrum of the reference and collected probe light back-scattering (mixed with noise), (e) the spectrum of the reference light, depleted probe light, c-SFG peak (red) and unfiltered i-SFG (red) background, (f) the narrowband optical filtering of i-SFG, (g) the homodyne beating frequency \(f_{HDD}\) between the DFG pump light and frequency shifted c-SFG.\label{schematic}} 
\end{figure*}
The probe and reference light of chaotic-QFC LiDAR is generated through difference frequency generation (DFG) between a single frequency pump (\(2\omg_0\)) and broadband amplified spontaneous emission (ASE, frequency higher than \(\omg_0\)).
In the limit of broadband DFG phase matching, the probe and reference light temporal amplitudes \(A_p, A_r\) (with carrier frequency \(\omg_0\) subtracted) are chaotic and complex conjugated to each other.
 Under the quasi-cw assumption, the probe and reference light can be modeled as stationary Gaussian random processes that are characterized by their correlation function \(f(t-t')\): 
\begin{gather}
	\avg{A_p(t)A_p^*(t')} = P_p f(t-t')\label{corr} \\
	A_r(t) = \sqrt{\frac{P_r}{P_p}} A_p^*(t)\label{conj}
\end{gather}
where \(P_p, P_r\) are the photon flux of the probe and reference light, respectively, and angled brackets stand for statistical averaging.
The chaotic and conjugated phases of the probe and reference light can be understood as classical time-frequency correlation (see the Discussion section for details).
Background noise light with flux \(P_n\) will naturally be uncorrelated with the reference light and is assumed to have the same temporal-spectral characteristics (correlation function) as probe light: 
\begin{gather}
	\avg{A_n(t)A_n^*(t')} = P_nf(t-t')\label{corrn} 
\end{gather}
For simplicity, the correlation function \(f\) is assumed to be Gaussian: 
\begin{gather}
	f(t-t')=\exp\left(-\frac{(t-t')^2\sigma^2}{2}\right)
\end{gather}
where \(\sigma\) is the common bandwidth of probe, reference, and noise light.
This condition is a worst-case scenario asymptotic case, where it is assumed that the noise involved is fully in-band, which cannot be removed through simple filtering.
In the LiDAR transceiver, probe light is mixed with strong background noise and collected by the telescope (Fig.\ref{setup}).\\

At the receiver section, the locally stored reference light can selectively convert probe light to another frequency via SFG, with little crosstalk from the noise light that has identical time-frequency properties.
To analyze this, consider first the small signal SFG regime, in which the SFG output amplitude is given by\cite{supp}:
\begin{gather}
	A_{SFG}(t) = \frac{\gamma}{\Delta\beta}\left(A_p(t)A_r(t)+A_n(t)A_r(t)\right)*\Pi\left(\frac{t}{\Delta\beta L}\right)
\end{gather}
where \(\gamma,L\) specifies the normalized nonlinearity and waveguide length and \(*\) stands for convolution.
The rectangle function \(\Pi(t)\) equals unity for \(-1/2< t \le 1/2\) and zero otherwise.
The inverse group velocity difference \(\Delta \beta\) is defined as the difference between the inverse group velocity of SFG and probe light.
\rev{The group velocity of noise and reference light is assumed to be the same as the probe light because of negligible dispersion around \(\omg_0\)\cite{reddy2013temporal}.
As a consequence, all frequency components of the probe, reference, and noise light are assumed to take part in the SFG process with the same efficiency.}  
Because the probe, reference, and noise light are all stationary random processes, so is the SFG output.
Therefore the SFG power spectral density \(S(\omg)\) is given by the Fourier transform of the correlation function: 
\begin{gather}
	S(\omg) =\frac{1}{2\pi}\mathcal{F}\avg{A_{SFG}(t)A^*_{SFG}(t+\tau)}_t \\
=\gamma^2L^2\text{sinc}^2\left(\frac{\Delta \beta L\omg}{2}\right)\{P_pP_r\delta(\omg)\nonumber\\
+\frac{(P_n+P_p/2)P_r}{2\sqrt{\pi}\sigma}\exp\left(-\frac{\omg^2}{4\sigma^2}\right)\}  \label{spectrum_expr}
\end{gather}
where \(\delta(\omg)\) is the delta function.
The two terms above can be understood as follow: the random, but correlated, phase of the reference and probe light cancel each other and result in a single-frequency coherent SFG (c-SFG) peak.
In contrast, the noise and reference light will only contribute to broadband incoherent SFG (i-SFG).
It is worth noting that the probe and reference light will also generate a small amount of i-SFG (as compared to c-SFG) in the absence of noise light.
This is because the inherent intensity fluctuation of the chaotic probe and reference light will create a chaotic component (i-SFG) of the SFG output.
Noise reduction of chaotic-QFC LiDAR is achieved by applying narrowband optical filtering to separate c-SFG from i-SFG.
The signal-to-noise ratio (SNR) of chaotic-QFC LiDAR is then given by the ratio between c-SFG and i-SFG power:   
\begin{gather}
	\text{SNR}_{QFC} = \frac{2\sqrt{\pi}\sigma}{BW}\frac{2P_p}{2P_n+P_p}\label{SNR}
\end{gather}
where \(BW\) is the filter (optical and electrical) bandwidth.
If the receiver does not have access to the correlation information (reference light), background noise will appear to be completely indistinguishable from the probe light.
Then the only useful information that can be extracted from collected light is the change of optical power due to target reflection.
The SNR of such direct detection is given by \(\text{SNR}_{DD}=P_p/(P_n+P_p)\), compared to which the SNR enhancement of chaotic-QFC LiDAR is given by:
	\begin{gather}
		\frac{SNR_{QFC}}{SNR_{DD}} = \frac{2\sqrt{\pi}\sigma}{BW}\label{ratio}
	\end{gather}
For example, if the probe spectrum is 7.5nm wide around 1560nm (in FWHM, obtainable with commercial filters for fiber optical communication) and the filter bandwidth is 10Hz, the SNR enhancement is around 111dB.\\

The chaotic-QFC LiDAR sensitivity is dictated by the c-SFG efficiency beyond the small signal SFG regime, which is defined as the probability of converting an incoming probe photon to a c-SFG photon.
To maximize the c-SFG efficiency \(\eta_c\), the i-SFG power needs to be minimized due to the energy conservation constraint.
This can be achieved with a narrow SFG phase matching bandwidth (long waveguide) or cavity resonance at the c-SFG frequency\cite{sensarn2009resonant}.
In the limit of negligible i-SFG power, the c-SFG efficiency can be analytically solved beyond the small signal SFG regime\cite{supp}:     
\begin{gather}
	\eta_{c} = \sin^2(\gamma\sqrt{ P_r}L)\exp\left(-\frac{\beta^2\Delta L^2\sigma^2}{2}\right)\label{eff_expr}
\end{gather}
where \(\Delta L\) is the relative distance between the probe and reference light.
In particular, for zero relative delay \(\Delta L=0\) and some finite reference power \(P_r\), the coherent conversion efficiency can reach 100\%.
Such conversion will also preserve the quantum properties of the input probe light because c-SFG is an intrinsically noiseless process that acts like a frequency domain beam-splitter\cite{kobayashi2016frequency}.
\rev{The oscillatory behavior of the SFG efficiency is because of the transition of SFG to DFG after the probe light is completely depleted.} For the general case of non-negligible i-SFG power, the c-SFG efficiency can be calculated through Monte Carlo simulation by drawing different samples of the chaotic probe and reference amplitudes and numerically solving the coupled mode equations.
Numerical results show that for a 5cm long waveguide with 0.09 group index difference (extracted from the second harmonic generation phase-matching bandwidth, for around 1560nm pump light) between SFG and probe/reference light, the maximal c-SFG efficiency can reach 92\%.
Unlike QI, the target distance for chaotic-QFC LiDAR does not need to be stabilized down to the sub-wavelength level even though a similar coherent receiver is used.
This is because the c-SFG power is not fast varying as a function of the probe light phase.
The distance of the target, however, can be determined by scanning the reference light delay and monitoring the c-SFG power.
The distance resolution is inversely proportional to the probe light bandwidth (\(\simeq\)100um for 7.5nm probe bandwidth).\\

The schematic of different stages of the LiDAR system and corresponding spectra of interacting light waves are summarized in Fig.
\ref{schematic} (a-g).
The probe light is created as a spectral mirror image (Fig.\ref{schematic} (b)) of the reference light (ASE) with conjugated phase, as per the energy conservation requirement of DFG.
Before SFG, the group velocity dispersion and relative delay of the probe and reference light need to be compensated, such that different frequency components of the probe and reference light contribute to c-SFG constructively.
The filtering of i-SFG consists of three stages: (1) the phase matching bandwidth (Fig.\ref{schematic} (e)) limits the i-SFG generation spectrum and (2) an optical bandpass filter (Fig.\ref{schematic} (f)) to optically reject most of the i-SFG power to prevent detector saturation and (3) a very narrowband electrical filter (integration time) of the phase-sensitive (balanced homodyne) detection (Fig.\ref{schematic} (g)) to maximize the noise rejection.
To avoid low-frequency technical noise, the frequency of the homodyne detection signal can be shifted to a non-dc frequency \(f_{HDD}\) by frequency shifting the probe light(Fig.\ref{schematic} (g)).

\section{result}
\begin{figure}[t]
	\centering
	\includegraphics[width=8cm]{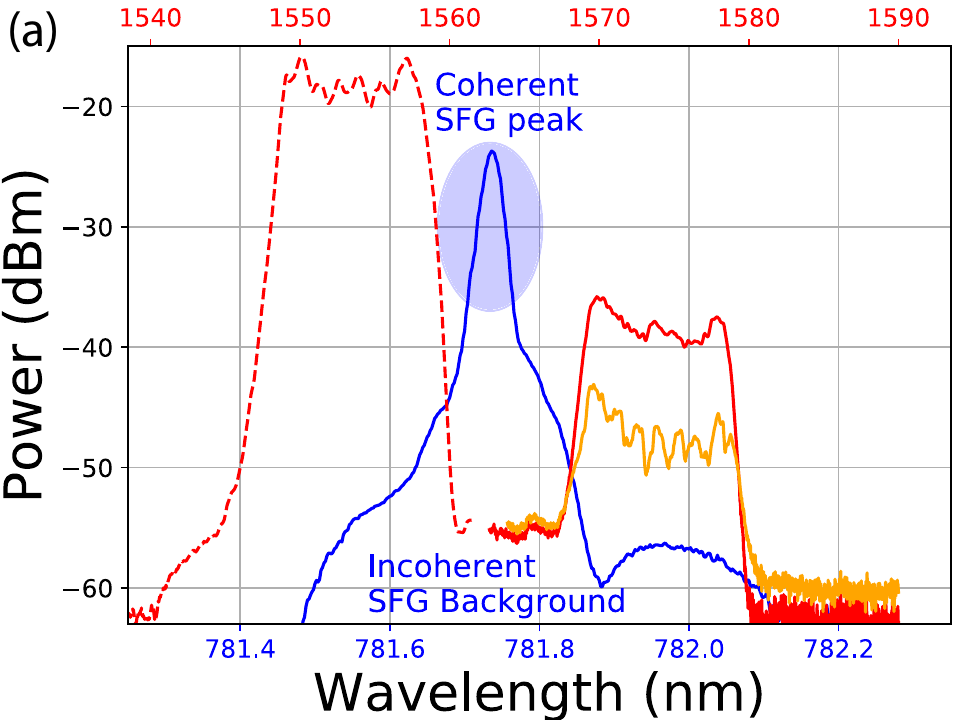}\\
	\hspace{0.2cm}
	\includegraphics[width=8cm]{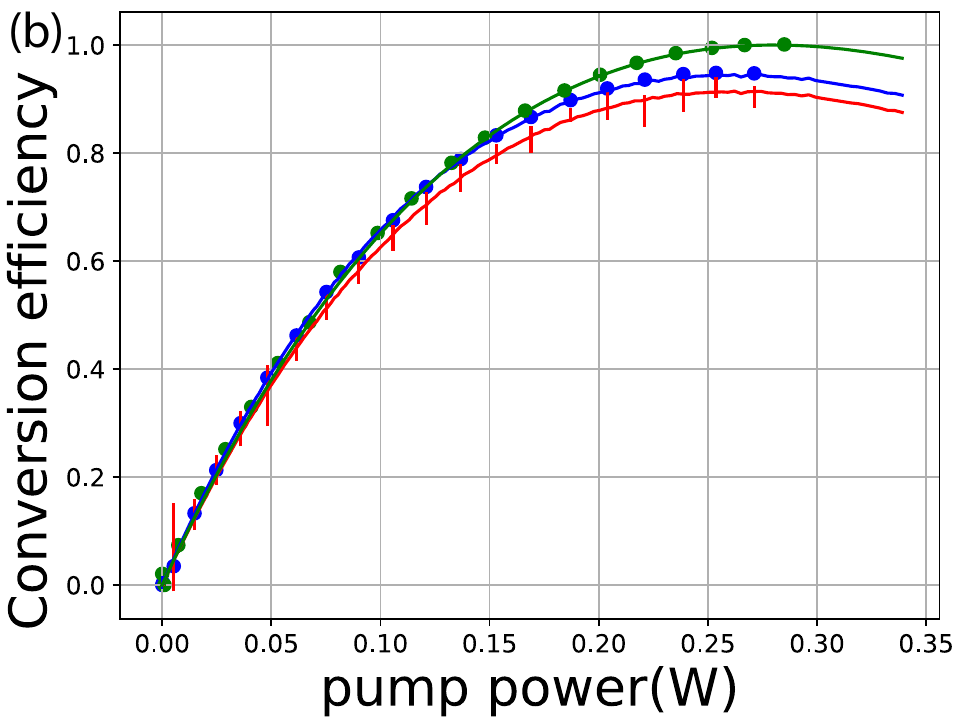}
	\caption{(a) the spectrum of reference light (dashed red), probe light before SFG (solid red), probe light after SFG (orange), and SFG light (blue).
The SFG spectrum (0.01nm resolution) consists of a c-SFG peak and broadband i-SFG background.
(b) the theoretically predicted (solid line) and experimentally measured (errorbar or dot) c-SFG efficiency (red) and total SFG efficiency (blue).
The SFG efficiency of the single-frequency probe and reference light is also plotted (green dot and curve) as a baseline.\label{eff} }
\end{figure}
\begin{figure}[t]
	\centering
	\includegraphics[width=7cm]{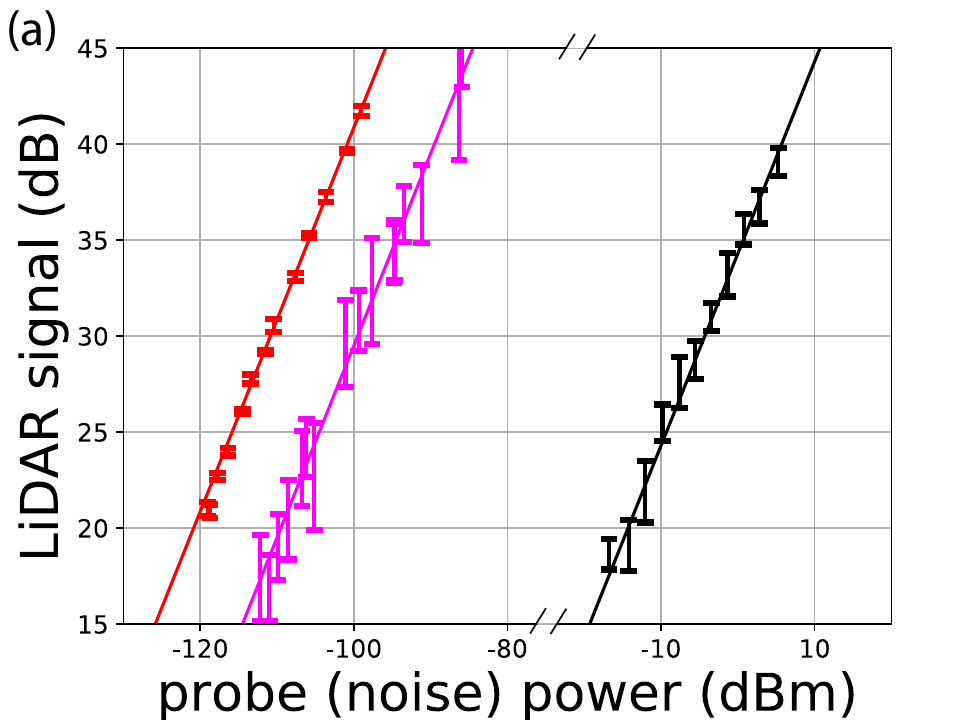}\\
	\hspace{0.2cm}
	\includegraphics[width=7cm]{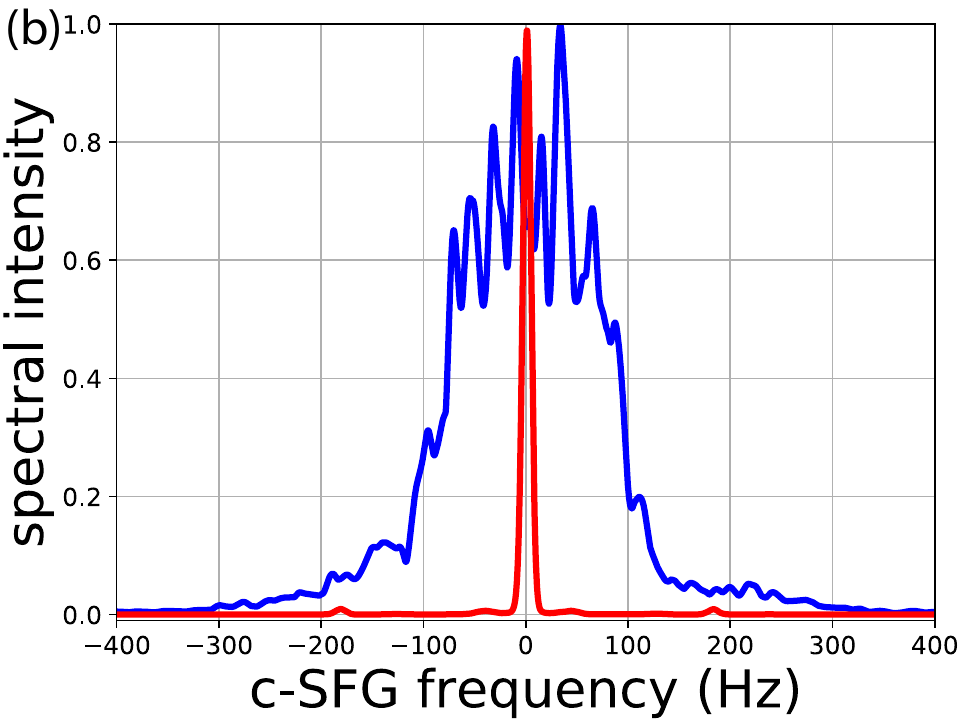}
	\caption{(a)the homodyne signal level (relative to shot noise level, 100ms integration time) for different probe and noise power.
Red: probe light back-reflected from a rough glass surface mounted on the optical table, magenta: probe light back-reflected from the quadcopter with phase noise induced by mechanical vibration, black: noise light.
The reference power is kept around 260mW to maximize the conversion efficiency.
(b) the ESA spectrum of the homodyne signal, when the stabilized ground glass (red) or quadcopter (blue), is used as the target.
\rev{Both spectra are normalized in the maximal spectral power level.} The difference in the bandwidths is due to mechanical vibration-induced broadening.
	\label{noise}}
\end{figure}
The LiDAR experimental setup is shown in Fig.\ref{setup}.
In the source section, the correlated probe and reference light (Fig.\ref{eff}(a)) is generated through DFG between the 781.7nm pump light and ASE light.
The spectrum of the ASE is limited to around 7.5nm in full-width half max (FWHM) by a tunable band pass filter.
A WDM filter and erbium-doped fiber amplifiers in C and L band are then used to separate and boost up the probe and reference light power.
To achieve the maximal c-SFG efficiency, the \rev{total group velocity dispersion (1.6ps/nm)} and path length difference of the probe and reference light is compensated by a programmable amplitude-phase filter (Finisar waveshaper 1000A) and an optical delay line.
The transceiver section consists of a probe light collimator, a homemade telescope, and a target object placed around 4 meters away.
Two different target objects are tested: a piece of grounded glass stabilized on the optical table and a quadcopter drone (with rough plastic surfaces) placed on top of a cart outside the optical table.
An adjustable level of noise power is simulated by mixing the telescope output with noise light that is obtained by tapping the probe amplifier output.
Since the tapped probe amplifier output is not correlated with reference light due to unbalanced delay, it can be regarded as uncorrelated background noise that has the same time-frequency properties as the probe light.
In the receiver section, collected probe light and reference light undergo SFG in a PPLN waveguide \rev{(HCP-RPE-5cm, details in coating, loss, and phase matching information can be found in \cite{supp})}, whose output first goes through a narrowband optical filter (diffraction grating) and then is phase-sensitively detected through balanced homodyne detection (with the local oscillator light tapped from the DFG pump light).
To measure the homodyne signal away from dc, the probe light is frequency-shifted by 2.1MHz through saw-tooth wave phase modulation (serrodyne).
The remaining non-converted probe and reference light are spectrally separated and the power is monitored.\\ 

The LiDAR sensitivity is benchmarked by the efficiencies of total SFG and c-SFG, which are defined as the probabilities of converting an incoming probe photon to an SFG (including both i-SFG and c-SFG) photon or a c-SFG photon.
The total SFG efficiency is determined from the depletion of probe power throughput when the reference light is turned on, as shown by the blue dots in Fig. \ref{eff}(b).
The proportion of c-SFG within total SFG is measured by comparing the phase-sensitive (balanced homodyne) detection and the intensity detection result of the total SFG.
Then the c-SFG efficiency can be calculated as the product of total SFG efficiency and c-SFG proportion, as shown by the red error bars in Fig. \ref{eff}(b).
To calibrate the measured c-SFG proportion to be independent of optical and detection efficiencies, a baseline measurement (Fig. \ref{eff}(b), green dots) is done with single-frequency probe and reference light, which only produces c-SFG.
The experimentally measured efficiencies of total SFG, c-SFG, and single frequency SFG are in agreement with the Monte Carlo simulation (solid lines in Fig.\ref{eff}(b))\cite{supp}.
The c-SFG efficiency with around 260mW input reference power reaches up to 90\%.
The measurement uncertainty is mainly due to the instability of the DFG pump laser's coherence.\\
The LiDAR noise resilience is benchmarked by comparing the homodyne signal level for different probe and noise powers (as measured at the output of the SFG waveguide).
It is worth emphasizing that the noise light has identical time-frequency properties as probe light and therefore cannot be separated from probe light via conventional time-frequency filtering.
Instead, the probe light is converted to c-SFG and separated from noise light induced i-SFG via narrowband optical and electrical filtering.
\rev{In the current experimental setup, the waveguide phase matching and diffraction grating contribute to around 97\% reduction of i-SFG (0.12nm filter bandwidth being applied on 3.7nm non-phase matched i-SFG bandwidth).
Additional optical filtering can be applied by adding a high-fineness Fabre-Perot filter.
} 
The narrowband electrical filtering is simulated by using 10Hz resolution bandwidth (100ms integration time) of an electrical spectrum analyzer (ESA).
As can be seen in Fig.\ref{noise}(a), to produce the same homodyne signal level, the noise power needs to be 107dB higher than probe light back-reflected from a stabilized target.
The reason for the measured 107dB noise rejection being lower than the theoretical prediction  (111dB from Eq.
\eqref{ratio}) can be attributed to many factors, including non-perfect frequency shift, non-Gaussian shaped spectra of probe and reference light, etc.
When the unstabilized quadcopter is used as the target object, back-reflected probe light (with the same flux as the stabilized target case) generated a weaker homodyne signal and provides less noise rejection (95dB).
This is because mechanical vibrations induce \(\simeq\)1kHz phase noise of the back-reflected probe light.
Such phase noise is transferred to the c-SFG spectrum  (Fig. \ref{noise}(b)).
As a result, the 10Hz filter also reduces the level of the measured homodyne signal due to over-filtering (10Hz\(<\)1000Hz).
\rev{It is worth noting that in the current setup, the nonzero DFG pump bandwidth (75KHz) does not spectrally broaden the homodyne signal.
This is because phase fluctuations of c-SFG and the local oscillator are identical and cancel each other in the homodyne detection.} \\
\begin{figure}[h]
	\centering
	\includegraphics[width=8cm]{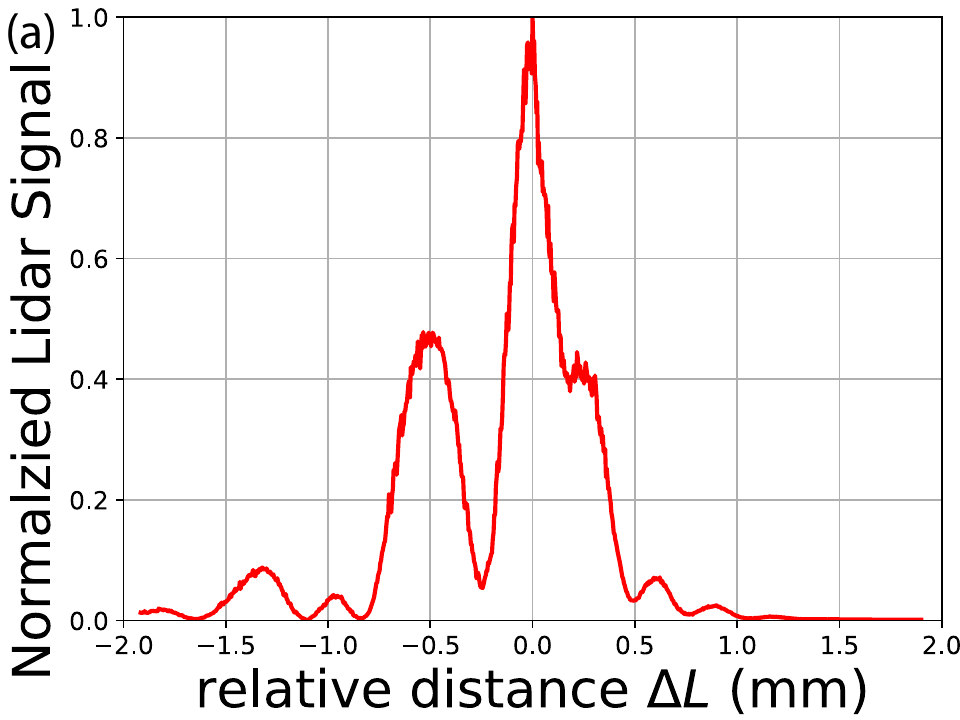}\\
	\hspace{0.2cm}
	\includegraphics[width=8cm]{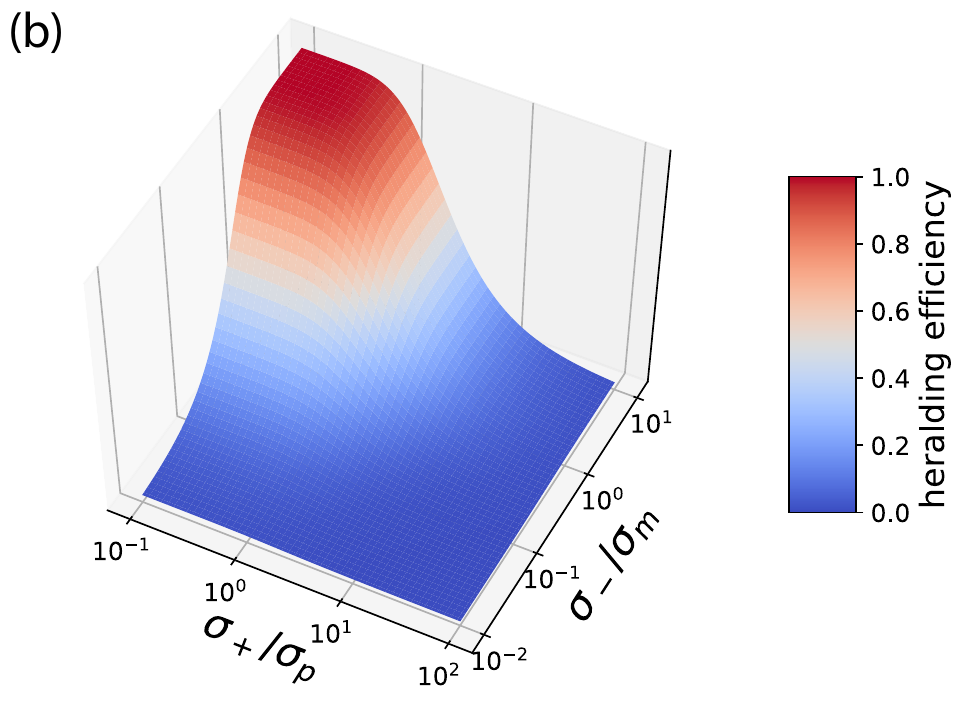}
	\caption{(a)The normalized homodyne signal as a function of reference light (relative) delay.
The resolution bandwidth is increased to 1KHz to reduce the impact of the Dopler effect on the homodyne signal level.
(b)the relationship between the probe mode heralding probability and the temporal duration \(\sigma_+\) and coherence time \(\sigma_-\) of the probe/reference CMs.\label{heralding}}
\end{figure}
To benchmark the ranging performance, the delay of the reference light is scanned with a homemade mechanical delay line to determine the target object(quadcopter) distance (Fig.\ref{heralding}(a)).
This is possible since c-SFG efficiency is non-negligible only if the probe and reference light have a balanced delay as can be seen from \eqref{eff_expr}.
\rev{When scanning, a Dopler frequency shift of homodyne signal that is proportional to the scanning speed is observed.} The ranging result shows multiple side peaks, \rev{ which is possibly caused by diffusive reflections of probe photons into different directions, resulting in different probe path lengths inside the telescope.
It is confirmed that when an object with specular reflection (mirror) is used as the target object, the ranging signal is a single peak localized in distance.
It is also worth mentioning that the speed and dynamic range of delay scanning in our current setup can potentially be improved with commercial solid-state optical switches, and fixed fiber optical delay lines.}

\section{Discussion}
\rev{The idea of utilizing sum frequency generation to analyze classical time-frequency correlation has been demonstrated in chirped pulse interferometry (CPI)\cite{kaltenbaek2008quantum,lavoie2009quantum}.
However, unlike in CPI where correlation is manually created by preparing laser pulses with opposite chirps, the correlation used chaotic-QFC LiDAR originates from the random and chaotic nature of broadband ASE light and the DFG energy conservation constraint, in a way that is similar to the generation of time-frequency entangled photon pairs.} This can be seen from Eq. \eqref{spectrum_expr} and \eqref{eff_expr} as follows: probe and reference light generate SFG mostly at a single frequency (c-SFG) despite both being broadband in the spectrum.
Also, c-SFG is efficient only if the probe and reference light has zero relative delay, despite both being stationary in time.
This is similar to the SFG process of time-frequency entangled photon pairs where two entangled photons must have close to zero relative delay and can only generate SFG photons at the sum frequency\cite{liu2020joint}.
Also, time-frequency entanglement and classical correlation are both generated via parametric down-conversion processes but with different input stimulation (vacuum versus ASE).
The essential difference between the classical correlation and entanglement is that: while entanglement can be simultaneously measured in time and frequency domain through joint measurement\cite{liu2020joint}, classical time and frequency correlation are actually the same correlation being represented in the time or frequency basis.
As a result, classical time-frequency correlations are subjected to the time-frequency uncertainty constraint.
For LiDAR, this implies that the noise rejection provided by the correlation is exactly half (in log scale) of that provided by entanglement.
Nevertheless, the classical nature of the correlation allows for probe and reference power beyond the single photon level and the option of classical amplification.
This is of paramount utility for practical applications where high probing power is required to extend the detection range beyond that of a proof-of-principle laboratory setup with single photon sources.
\\

The noise light for chaotic-QFC LiDAR is assumed to be statistically identical to probe light since out-of-band noise (in either time or frequency domain) can be filtered without impacting the LiDAR performance.
Also, from a practical point of view, indistinguishable noise light simulates active jamming attacks or unintentional interference (e.g. identical adjacent LiDARs operating at the same time), which are important safety concerns \rev{that degrade the sensitivity and increase the error rates of LiDARs\cite{hwang2020mutual}.
The conventional approach to this problem is to generate spread spectrum probe light and perform correlation analysis in the radio frequency domain, i.e. chaotic LiDARs based on chaotic lasers\cite{lin2004chaotic} and random modulation\cite{hwang2020mutual,feng2021fpga}.
In comparison, the time-frequency correlation used in chaotic-QFC LiDAR can be considered a spread spectrum technique in the optical domain that has two vital differences.
First, the optical time-frequency correlation is much broader in bandwidth than the RF domain spread spectrum.
This translates to orders of magnitude superior noise suppression and ranging resolution.
Secondly, the time-frequency correlation is optically analyzed through \(\chi^{(2)}\) nonlinear process, and doing so also avoids fundamental limitations of electrical domain measurement and signal processing such as detector saturation, electrical interference, finite sampling rate, digitization noise, and imperfect common mode rejection ratio of balanced photodetection, etc.
A brief review and comparison of different types of LiDAR-based on classical correlation are given in \cite{supp,lin2004chaotic,hwang2020mutual,feng2021fpga,shahverdi2018mode,hwang2019study}}.

In chaotic-QFC LiDAR, probe light with a specific chaotic amplitude is separated from uncorrelated noise light via c-SFG.
\rev{It is interesting to note that a one-to-one correspondence can be made between this and the process of QFC: in QFC, a quantum state of light (corresponding to the LiDAR probe light) in a particular time-frequency mode (probe mode) is coherently converted to another mode (the c-SFG mode) and separated from other quantum states (noise light), through nonlinear interaction with a strong pump light (reference light).
Also, the high-efficiency nature of chaotic-QFC preserves the quantum properties\cite{eckstein2011quantum} of the input probe light like other QFC protocols.
For this reason, chaotic-QFC can be considered as a variant of QFC that is equipped with chaotic modes (CMs) defined by chaotic temporal amplitudes.
Nevertheless, there are important differences between chaotic-QFC from conventional QFC based on Hermite-Gaussian modes, in terms of orthogonality and statistical equivalence\cite{supp}.
These two differences have important implications on QFC performance metrics, which are summarised as follows.}\\

\rev{First, different CMs are approximately orthogonal due to their independent chaotic nature.
Nevertheless, if each CM has a sufficiently large time-bandwidth product, such approximate orthogonality is close to exact and can simultaneously provide high conversion efficiency and selectivity (the conversion efficiency ratio of the target mode versus other crosstalk modes\cite{reddy2018high,huang2013mode}).
To give a concrete example, consider a target CM with 100ns duration and 1THz bandwidth.
Since the 100ns pulse width is much longer than the nonlinear interaction time, the previous quasi-cw modeling of chaotic-QFC still applies.
Therefore the c-SFG efficiency will still reach over 90\% as has been experimentally observed (can be further improved with narrower SFG phase matching).
The generated c-SFG is a 100ns pulse with a single carrier frequency, whereas i-SFG from other crosstalk modes is evenly distributed around 1THz bandwidth.
As a result, by using a \(\simeq\) 10MHz optical filter, the i-SFG power can be rejected by around \(\simeq 1- 10^{5}\).
In comparison to this, state of art QFC based on cascaded nonlinear conversion has been shown to provide \(\simeq 90\%\) of crosstalk mode rejection\cite{reddy2018high}.}\\

Another important difference of CMs is that they have similar \textit{noise-like} time-frequency characteristics, whereas Hermite Gaussian modes of different orders have substantially different time-frequency amplitudes.
Therefore it is much easier for chaotic-QFC to multiplex many CMs in parallel to span a large Hilbert space for high-dimensional quantum information processing.
This can be carried out, for example, by taking different temporal segments of an ASE source, which can be done with integrated optical delay lines.
An additional advantage of chaotic-QFC as compared to Hermite-Gaussian mode based QFC is that the generated SFG is in the same time-frequency mode (c-SFG) regardless of the input CMs.
This makes the subsequent quantum processing of SFG light easier to standardize.
Moreover, this also allows for the interference of c-SFG from different probe modes.
When combined with the recently developed multiple output QFC technique \cite{serino2022realization,kruse2023pulsed}, the chaotic-QFC technique can be useful for frequency domain Boson sampling\cite{joshi2017frequency}.
The superior performance of chaotic-QFC does, however, come at a price of reduced channel capacity: to ensure approximate orthogonality, the number of CMs must be much lower than the total time-bandwidth product of the channel to ensure sufficiently low mutual overlap.\\

\rev{It is worth emphasizing that chaotic-QFC, like other QFC protocols, does not generate quantum states of light.
Instead, it serves as a means to manipulate quantum light that already exists in CMs through SFG.
Therefore, the question remaining is whether quantum states of light can be generated in CMs by some other nonlinear processes in the first place.
After all, a quantum light-generating process such as spontaneously parametric down-conversion (SPDC), does favor an intrinsic set of modes (e.g. Schmidt modes \cite{ekert1995entangled} and Bloch-Messiah modes\cite{wasilewski2006pulsed}) depending on the nonlinear medium and pumping.} What we shall show next is that phase conjugated pairs of CMs are \textit{approximately} intrinsic (Schmidt) mode pairs of SPDC processes that have narrow band pump and broadband phasematching.
For such SPDC processes, the joint temporal amplitude (JTA) can be approximated by a delta function \(\delta(t_p-t_r)\), whose Schmidt decomposition is \textit{not} unique:  
\newcommand{\ket}[1]{|#1\rangle}
\begin{gather}
	\ket{\text{pair}} = \sum_k\int dt_p f_k(t_p)\ket{t_p} \int dt_r f_k^*(t_r)\ket{t_r}
\end{gather}
where \(\{f_k(\omg)\}\) is an arbitrary basis of square normalized functions and single photon probe and reference states at time \(t_p,t_r\) are denoted as \(\ket{t_p}, \ket{t_r}\), respectively.
\rev{The arbitrariness of \(f_k(t)\) implies if a photon is created in a  CM with amplitude \(A_p(t)\), its twin photon will also be created in the conjugated CM with amplitude  \(A_r(t)=A_p^*(t)\).}\\

\rev{To verify the validity of the above argument beyond the limiting case discussed above, consider the following photon pair state with a Gaussian JTA:}
\begin{gather}
	\ket{\text{pair}} = \iint dt_pdt_r\nonumber\\
	\sqrt{ \frac{1}{2\pi\sigma_p\sigma_m} \exp\left(-\frac{(t_p-t_r)^2}{4\sigma_m^2}-\frac{(t_p+t_r)^2}{4\sigma_p^2}\right) }\ket{t_p}\ket{t_r}\\
\end{gather}
\rev{
where \(\sigma_m,\sigma_p\) are the temporal correlation and duration of SPDC photon pairs.
Then consider a pair of CMs whose temporal amplitude \(A_p(t),A_r(t)\) are specified by the correlation function:}
\begin{gather}
	\avg{A_p(t)A_p^*(t')} = \frac{1}{\sqrt{2\pi}\sigma_-}\exp\left(-\frac{(t-t')^2}{8\sigma_-^2}-\frac{(t+t')^2}{8\sigma_+^2}\right)\\
	A_r(t) =  A_p^*(t)
\end{gather}
\rev{where \(\sigma_-,\sigma_+\) are the temporal correlation and duration of two CMs.
 Then fidelity of these two CMs as a pair of Schmidt modes can be benchmarked by the heralding efficiency in mode \(A_r(t)\) when a photon in \(A_p(t)\) is detected.
As can be seen in Fig. \ref{heralding}, near unity heralding efficiency can be achieved when the coherence time \(\sigma_-\) of the CMs is longer than the SPDC temporal correlation time while the temporal duration \(\sigma_+\) of the CMs needs to be shorter than the SPDC photon temporal duration \(\sigma_p\).
}

\section{conclusion}
In summary, we proposed a quantum-inspired LiDAR prototype based on coherent measurement of classical time-frequency correlation.
It retains the high noise resilience advantage (\(>\)100dB rejection of indistinguishable noise) of quantum LiDARs while still allowing for high-power classical sources and single photon sensitivity to extend the detection range.
Its principle also resembles chaotic LiDAR but the implementation with coherent optical processing avoids fundamental limitations of electrical domain detection and signal processing.
As such, the LiDAR prototype we demonstrate here combines both practical implementation and substantial performance enhancement.
We expected it to become a useful tool in the near future for real-world LiDAR application that requires high rejection of crosstalk and noise jamming.
The chaotic mode conversion technique that is derived from the LiDAR receiver, can also be applied in quantum information applications and provide performance enhancement as compared to pulse-based quantum frequency conversion.
Its advantage of high efficiency and selectivity can be useful for high dimensional quantum information processing applications such as Boson sampling.

\section*{Data Availability}
The authors declare that all necessary data to interpret, verify and extend the reported result have been provided in the main text and the supplementary information.
\section*{Code Availability}
The algorithm for the Monte-Carlo SFG simulation is described in the supplementary information and the code is available upon requests.
\bibliographystyle{unsrt}
\bibliography{references.bib}

\end{document}